\documentclass[AMA,Times1COL]{./WileyNJDv5_Template/WileyNJDv5}
\usepackage{import}
\usepackage[utf8]{inputenc}
\usepackage{graphicx}
\usepackage{caption}
\usepackage{booktabs}
\usepackage{subcaption}
\usepackage{comment}
\usepackage{hyperref}
\usepackage{svg}
\usepackage[section]{placeins}
\svgpath{{Figures/}}
\graphicspath{{Figures/}}

\articletype{Research Article}%

\received{Date Month Year}
\revised{Date Month Year}
\accepted{Date Month Year}
\journal{Journal}
\volume{00}
\copyyear{2023}
\startpage{1}

\begin{document}

\title{Breamy: An augmented reality mHealth prototype for surgical decision-making in breast cancer}

\author[1]{Niki Najafi}
\author[2]{Miranda Addie}
\author[3,4]{Sarkis Meterissian}
\author[1]{Marta Kersten-Oertel}

\authormark{Najafi \textsc{et al.}}
\titlemark{Breamy: An augmented reality mHealth prototype for surgical decision-making in breast cancer}

\address[1]{\orgdiv{Applied Perception Lab, Department of Computer Science and Software Engineering}, \orgname{Concordia University}, \orgaddress{\state{Québec}, \country{Canada}}}

\address[2]{\orgdiv{Experimental Surgery}, \orgname{Mcgill University}, \orgaddress{\state{Québec}, \country{Canada}}}

\address[3]{\orgdiv{Breast Center}, \orgname{McGill University Health Centre}, \orgaddress{\state{Québec}, \country{Canada}}}

\address[4] {\orgdiv{Department of Surgery, Faculty of Medicine}, \orgname McGill University, \orgaddress{\state{Québec}, \country{Canada}}}

\corres{Corresponding author Niki Najafi,  \email{niki.najafi@mail.concordia.ca}}
\begin{comment}
\presentaddress{This is sample for present address text this is sample for present address text.}
\end{comment}

%\fundingInfo{Text}
%\JELinfo{ejlje}

\abstract[Abstract]{In 2020, according to WHO, breast cancer affected 2.3 million women worldwide, resulting in 685,000 fatalities. By the end of the year, approximately 7.8 million women worldwide had survived their breast cancer making it the most widespread form of cancer globally \cite{whoBreastCancer}. Surgical treatment decisions, including choosing between oncoplastic options, often require quick decision-making within an 8-week time frame. However, many women lack the necessary knowledge and preparation for making such complex informed decisions. Anxiety and unsatisfactory outcomes can result from inadequate decision-making processes, leading to complications and the need for revision surgeries. Shared decision-making and personalized decision aids have shown positive effects on patient satisfaction and treatment outcomes. This paper introduces Breamy, a prototype mobile health (mHealth) application that utilizes augmented reality (AR) technology to assist breast cancer patients in making informed decisions. The app provides 3D visualizations of different oncoplastic procedures, aiming to improve confidence in surgical decision-making, reduce decisional regret, and enhance patient well-being after surgery. To determine the perception of the usefulness of Breamy, we collected data from 166 participants through an online survey. The results suggest that Breamy has the potential to reduce patient's anxiety levels and assist them during the decision-making process.}

\keywords{Augmented Reality, Breast cancer, Visualization, Decision-making, Surgical shared decision-making}

\jnlcitation{\cname{%
\author{Taylor M.},
\author{Lauritzen P},
\author{Erath C}, and
\author{Mittal R}}.
\ctitle{On simplifying ‘incremental remap’-based transport schemes.} \cjournal{\it J Comput Phys.} \cvol{2021;00(00):1--18}.}

\maketitle

\renewcommand\thefootnote{}
\footnotetext{}
\renewcommand\thefootnote{\fnsymbol{footnote}}

\section{Introduction}\label{sec1}
A 2020 study showed that breast cancer represents 1 out of every 8 cancer diagnosis, resulting in a collective count of 2.3 million new occurrences among both males and females \cite{sung2021global}. Approximately 685,000 women lost their lives to breast cancer in 2020, constituting around 16\% or 1 in every 6 fatalities caused by cancer among women \cite{whoBreastCancer}\cite{arnold2022current}.

 In the case where surgical treatment is required, breast cancer patients often rely on their physicians to decide between the various surgical and reconstructive options. These may include breast reconstruction, nipple reconstruction, etc. The time available for women to make their decision is ideally limited to 8 weeks, as this improves their chances of survival\cite{wiener2023reexamining}. However, many women, lack adequate knowledge and decisional preparation to make an informed decision within the available short time frame ~\cite{manne2016attitudes}. According to a 2019 research study, over 20\% of breast reconstructions fail to meet patient satisfaction, necessitating a subsequent revision surgery to address the unsatisfactory outcomes of the initial procedure \cite{lee2019factors}.  

% In addition, some centres lack a structured framework to assist women in decision-making~\cite{wiener2023reexamining}. Research has shown that anxiety can lead to lower satisfaction with the outcomes of breast reconstruction\cite{dasari2015rise}. Patients who experience anxiety before the procedure tend to have higher rates of complications, longer hospital stays, and increased costs following mastectomy\cite{rosenberg2015local}\cite{momoh2017tradeoffs}. 
\begin{comment}
Additionally, patients treated in public hospitals experienced a longer average time before undergoing revision surgery compared to those in private hospitals (26 months versus 12 months). This difference is likely due to resource constraints within the public hospital system\cite{finlay2021long}.
\end{comment}
Before surgery, patients meet with their plastic and breast surgeons who explain their treatment options using both spoken and written explanations, accompanied by pictures, including breast-conserving surgery or mastectomy. Women who are undergoing breast reconstruction will also often receive photographs of other women who have undergone similar procedures to better understand post-operative outcomes. However, some women believe that viewing these postoperative images does not always provide them with a clear understanding of their own potential appearance, and they have a hard time looking at the images \cite{godden2021three}. Studies have also shown that patients prefer making the decisions for themselves as they tend to experience higher satisfaction, better adherence to treatment, and improved quality of life compared to decisions made on their behalf \cite{murray2015satisfaction}. Furthermore, surgical shared decision-making (between patient and clinician) lowers decisional conflicts, anxiety, and the re-occurrence of surgical procedures. In addition, it improves knowledge retention, decisional satisfaction, and increased physician trust \cite{niburski2020shared}.

This paper introduces \emph{Breamy} a prototype application that uses augmented reality (AR) technology to assist breast cancer patients in making informed decisions. The app uses AR to project different oncoplastic procedures (e.g. breast reconstruction, nipple reconstruction, scaring possibilities, going flat, etc.) after mastectomy onto the patients body. Our aim is to provide patients and clinicians with a tool to improve confidence in surgical decision-making, lower decisional regret, and improve patient's quality of life after surgery. As a result, we hypothesize that fewer revision surgeries will be needed, which can lessen the burden on both patients and the healthcare system. 

\section{Related Works}\label{sec2}

Decision aids (DAs) aim to support patients by providing them with more information about medical procedures, helping them clarify their values, and involving them more actively in making decisions \cite{schonberg2014development}. Many studies have shown that the use of DAs improves medical decision-making \cite{o1999decision} \cite{stacey2017decision}. One type of decisional aid is simulation. The goal of simulation is to improve communication during the preoperative planning stage of surgery and to close the gap between patients' perceptions and their expectations. In breast cancer, 3D simulation has proven to be an effective implant selection tool \cite{hoffman2010decision}\cite{schonberg2006decision}\cite{torke2013older}\cite{o1999decision}, as well as a method of providing personalized healthcare that is commonly utilized to aid patients in making decisions \cite{lee2013time} \cite{crivellari2000burdens}. A recent study found that women who saw an individualized 3D simulation of their potential aesthetic outcome for breast-conserving treatment (BCT) were more confident heading into surgery than those who were given 2D images of other women\cite{godden2021three}. 

Augmented reality has been used in surgical planning and decision-making in a number of domains. Researchers have developed a markerless AR protocol to aid in the process of harvesting a skin paddle during an osteomyocutaneous fibular flap procedure. The authors employed the ``guideview" technology to initiate real-time tracking of the 3D printed phantom leg~\cite{cercenelli2022augmented}. Another study proved the usefulness AR to guide nerve sparing (NS) during robot-assisted radical prostatectomy (RARP) \cite{schiavina2021real}. For preoperative planning in maxillofacial surgery, surgeons have manipulated 3D cast models created from CT images\cite{fushima2016mixed}. In total knee arthroplasty, surgeons used augmented reality to plan cutting guides to do resections \cite{fucentese2021novel}.

For breast cancer specifically, Amini \emph{et al.}\cite{amini2022prototype} looked at reducing revision surgeries in breast cancer by improving implant decision-making. They designed a prototype pipeline for choosing the most appropriate implant to ensure the most natural breast shape. This was done using finite element modeling (FEM) of 3D patient models generated from pre-surgery MRIs. In a similar work, Amini \emph{et al.}\cite{amini2019augmented} developed an augmented reality application using the HoloLens that allows surgeons to go through various implant sizes (projected on a patient) in order to choose the most appropriate shape in single mastectomy (where the goal is to reconstruct one breast to look they as the healthy breast). The results suggest that computerized decision-making tools in breast reconstruction surgery can improve clinical decision-making and reduce revision surgeries.

The Vectra XT is a commercial device that captures (3D) photographic images and has the potential to aid in pre-operative planning and serve as a means of evaluating aesthetic breast reconstruction outcomes \cite{godden2021three}. The typical cost of this equipment is around \$15,000 \cite{bimedisCANFIELDVECTRA}. Breamy's proof of concept mobile health (mHealth) app aims to provide a similar solution in terms of visualizing multiple oncoplastic alternatives with the use of augmented reality (AR). mHealth apps aim to improve health outcomes and reduce costs by giving patients more information and connecting them with healthcare providers.  mHealth, applications are increasingly being used to provide personalized and more accessible care. In a recent review, out of the 69 breast cancer-related mHealth apps examined, the primary feature found in most of them was providing information on the early detection of breast cancer\cite{yang2023mobile}. To the best of our knowledge, no breast cancer mHealth applications provide an augmented reality surgical decision-making tool. With the introduction of Breamy, a new mobile AR app that offers customizable 3D breast models, we aim to offer a simulation of oncoplastic results in a cost-effective and easily accessible way, benefiting a wider population.

%A study found that patients treated in public hospitals had to wait longer for follow-up surgeries compared to those in private hospitals (26 months versus 12 months), likely because public hospitals have fewer resources \cite{finlay2021long}. However, with the help of Breamy, we are aiming to make resources more accessible to patients and public hospitals.
%More than half of all mobile phone users in the United States downloaded health-related applications in 2020, and two-thirds of those users said that doing so had improved their health \cite{krebs2015health}\cite{yang2023mobile}.

%The quality of these mHealth applications has been shown to be impacted by the engagement they provide, functionality, aesthetics, information, as well as other subjective features\cite{stoyanov2015mobile}\cite{terhorst2020validation}\cite{yang2023mobile}. 
%Moreover, breast cancer-focused mHealth apps that offer personalized data tend to offer higher quality. 

\section{Methodology}
In a recent paper~\cite{reyes2022user} on designing surgical systems, the authors argue that many research proof-of-concept systems are not translated into clinical practice due to a lack of multidisciplinary teams which include clinicians and patients, and a mismatch between technological and human needs. The paper suggests that a user-centered design (UCD) approach which focuses primarily on making systems more usable by prioritizing users, their activities, and contexts during each phase of a project, could ensure that surgical technologies are better translated into clinical contexts. UCD is an iterative design process that typically involves four steps: understanding the context of use through research and analysis, specifying user requirements and designing solutions, evaluating designs and implementing solutions. We used a user-centered design (UCD) approach with a multidisciplinary team of surgeons, clinical researchers, computer scientists, and patients to develop the Breamy prototype. We go through the steps of the process below.

\subsection{Research and Analysis}
To determine the most important needs for our application, a preliminary literature review, analysis of existing solutions, and interviews with patients and surgeons were conducted.  We identified the final needs and user requirements by evaluating various factors including "patient impact", "feasibility", and "potential to add value". Based on our research, we determined that there was a significant need to provide breast cancer patients undergoing mastectomy with personalized oncoplastic (e.g. breast reconstruction/cosmetic options) information to facilitate surgical decision-making and in turn increase their quality of life. Specifically, we found that patients with breast cancer wanted to have better knowledge of the disease, to better understand treatment options and prepare for treatment, to be able to better visualize body image changes, to enable self-management throughout their patient journey, and to have better support mental support \cite{alba2013adapting}\cite{lei2011informational}\cite{jeang2015process}\cite{hou2020development}. %\textcolor{blue}{The AR module of Breamy described in this paper is focused on enabling woman to better understand treatment options and visualize various body image changes that would improve decision making and satisfaction in surgical outcomes.}

\subsection{Design}
 %Motivators are crucial factors in encouraging breast cancer patients in utilizing the mHealth applications\cite{harder2017user}\cite{napoles2019feasibility}\cite{min2014daily}\cite{gehrke2018development}\cite{zhu2017development}\cite{lozano2019association}. Motivators include targeted suggestions\cite{gehrke2018development}\cite{lozano2019association}, professional guidance \cite{lozano2019association}, vivid interfaces \cite{min2014daily}, and informational and emotional support from peers or other significant people \cite{zhu2017development}\cite{lozano2019association} \cite{cai2021mobile}. Therefore, we designed Breamy, to have an appealing interface, guidance through an AR feature that allows visualizing various oncoplastic features, a knowledge repository to aid in informed decision making including information about post-surgery results, and a breast cancer community/support groups (see Figure \ref{fig:UI_design}). 

\begin{figure} [h!]
  \centering
  \begin{subfigure}[b]{0.34
  \textwidth}
    \includegraphics[width=\textwidth]{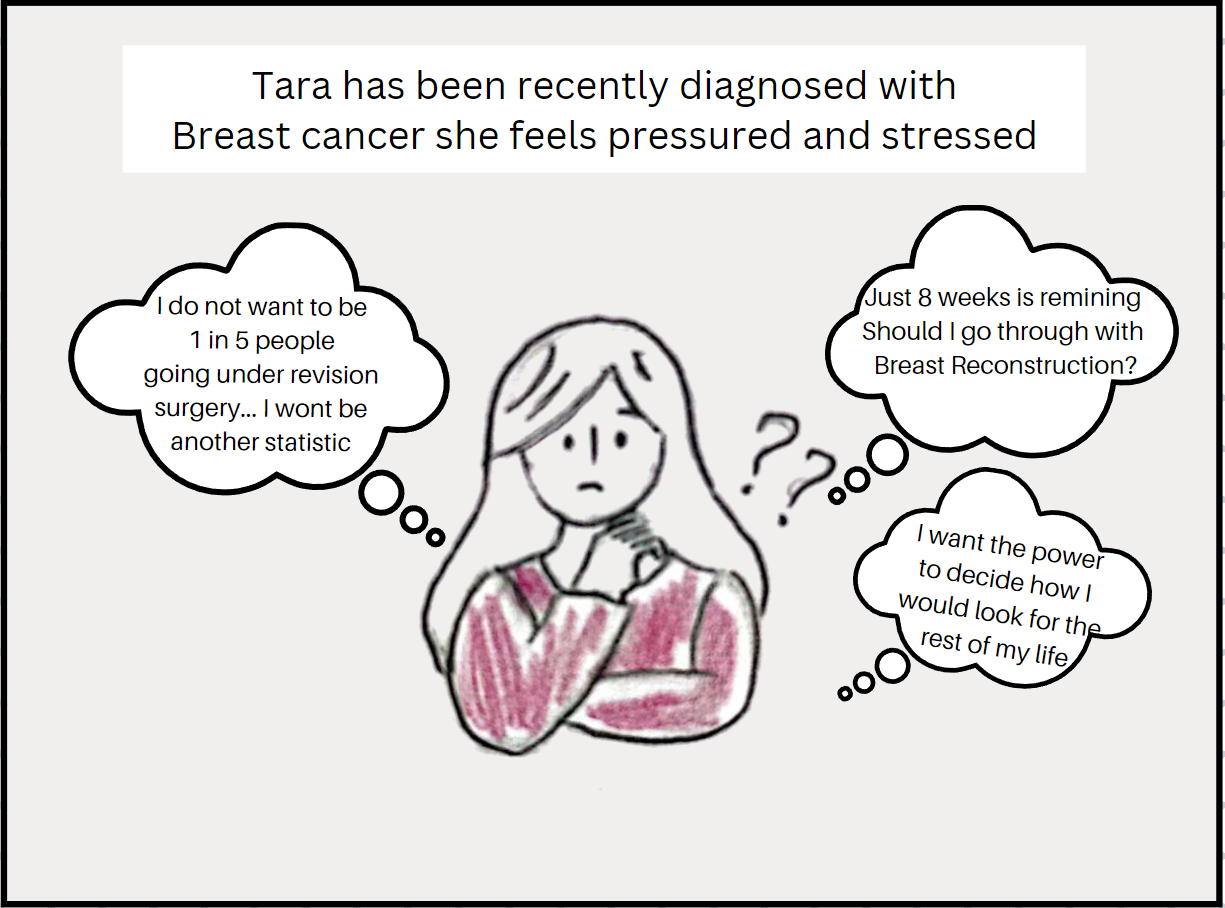}
    \caption{}
    \label{fig:image1}
  \end{subfigure}
  \hfill
  \begin{subfigure}[b]{0.322\textwidth}
    \includegraphics[width=\textwidth]{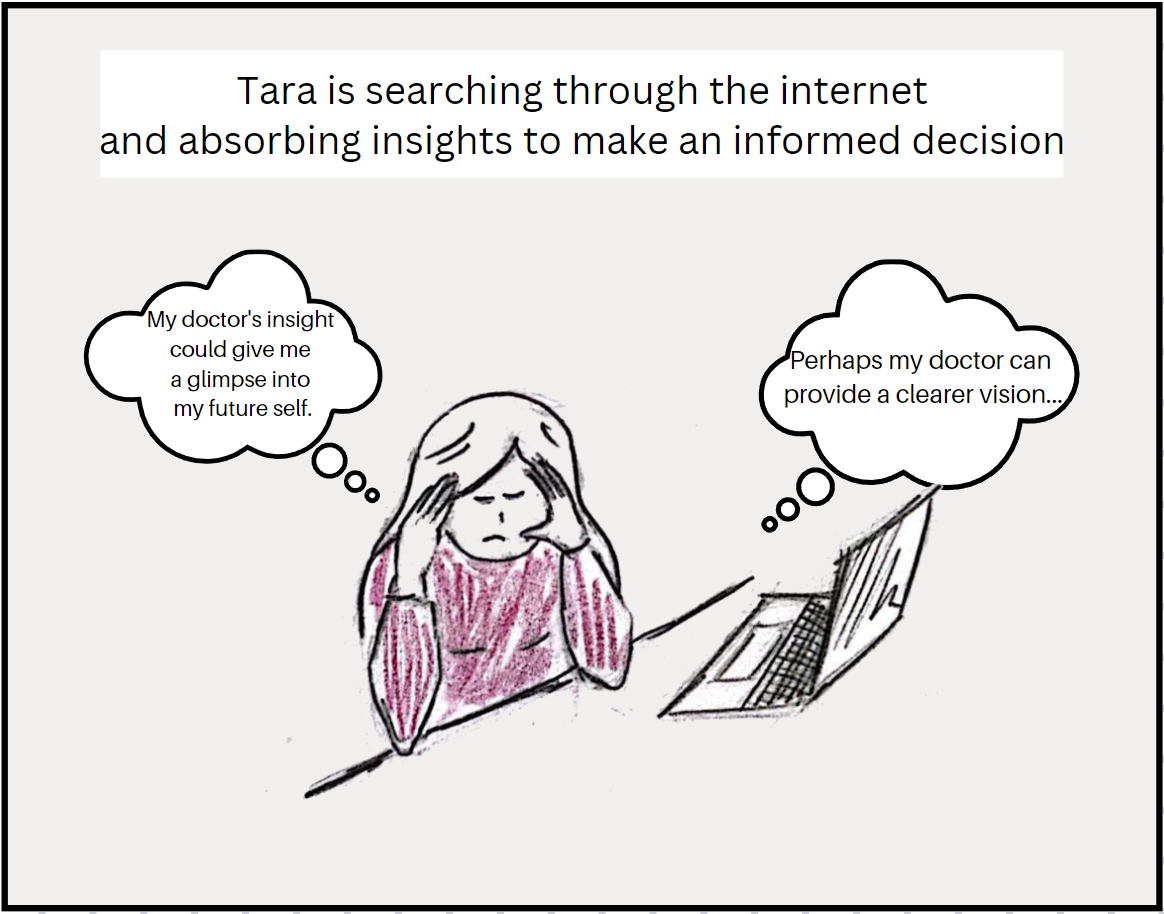}
 %   \caption{By exploring various oncoplastic options with AR, users will have the opportunity to visualize their potential appearance with or without clothing.}
 \caption{}
    \label{fig:3D visualization}
    \end{subfigure}
  \hfill
  \begin{subfigure}[b]{0.307\textwidth}
    \includegraphics[width=\textwidth]{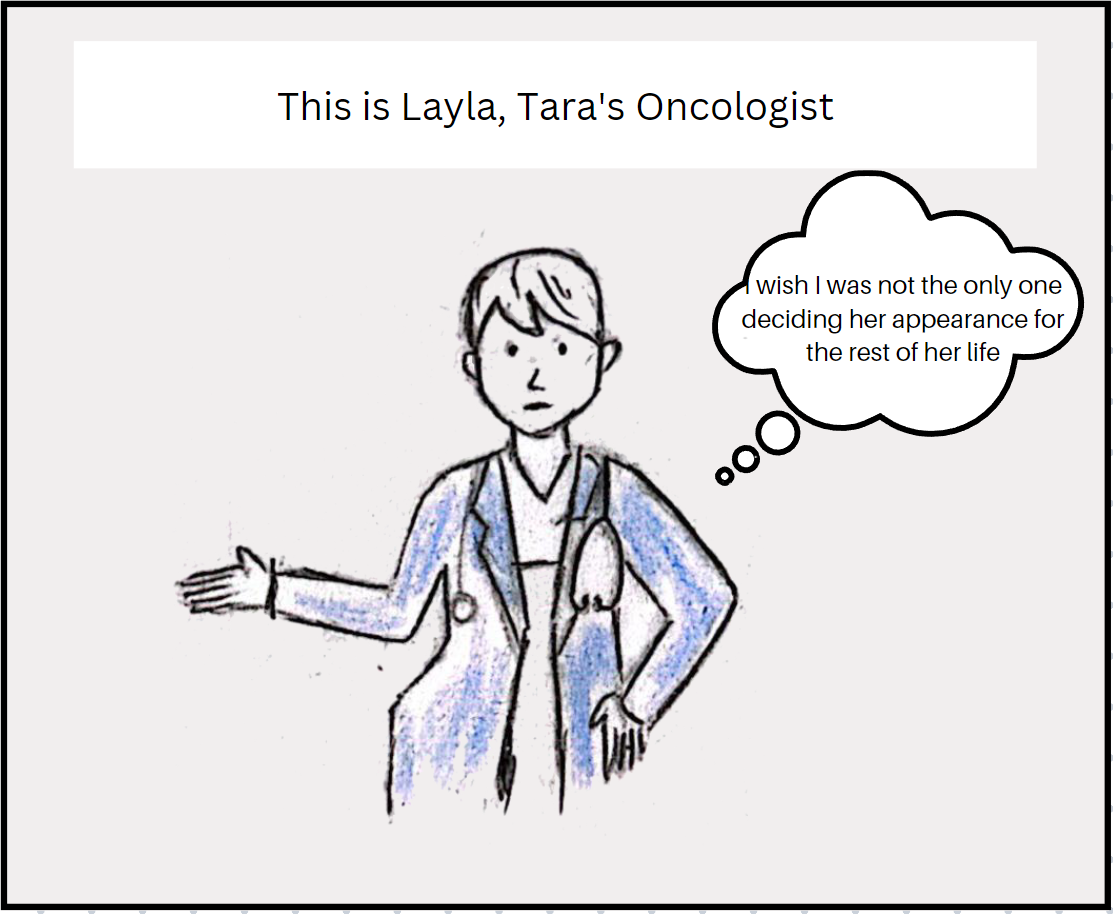}
    \caption{}
    \label{fig:3D visualization}
    \end{subfigure}
  \hfill
  \begin{subfigure}[b]{0.34\textwidth}
    \includegraphics[width=\textwidth]{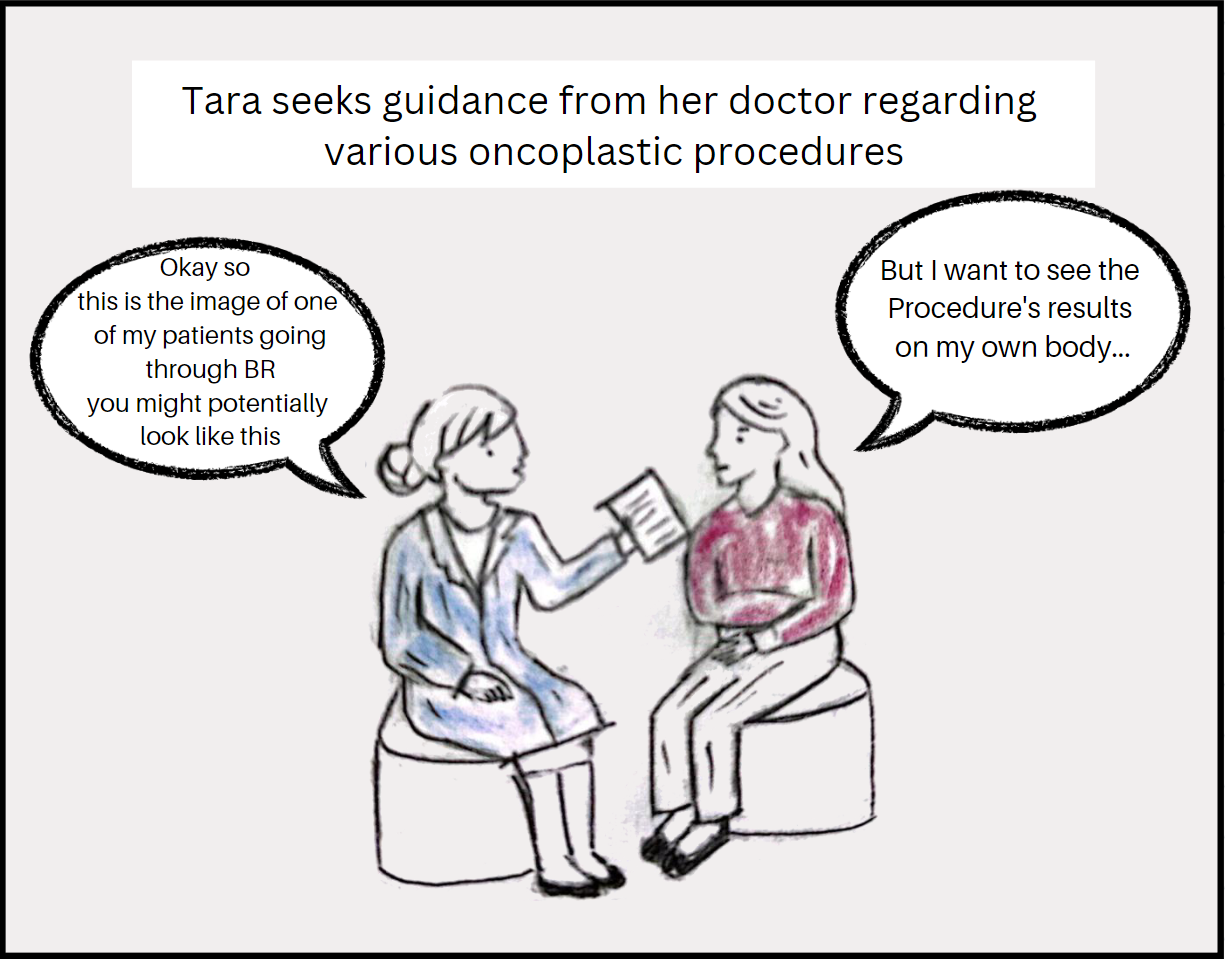}
    \caption{}
    \label{fig:3D visualization}
    \end{subfigure}
  \hfill
  \begin{subfigure}[b]{0.322\textwidth}
    \includegraphics[width=\textwidth]{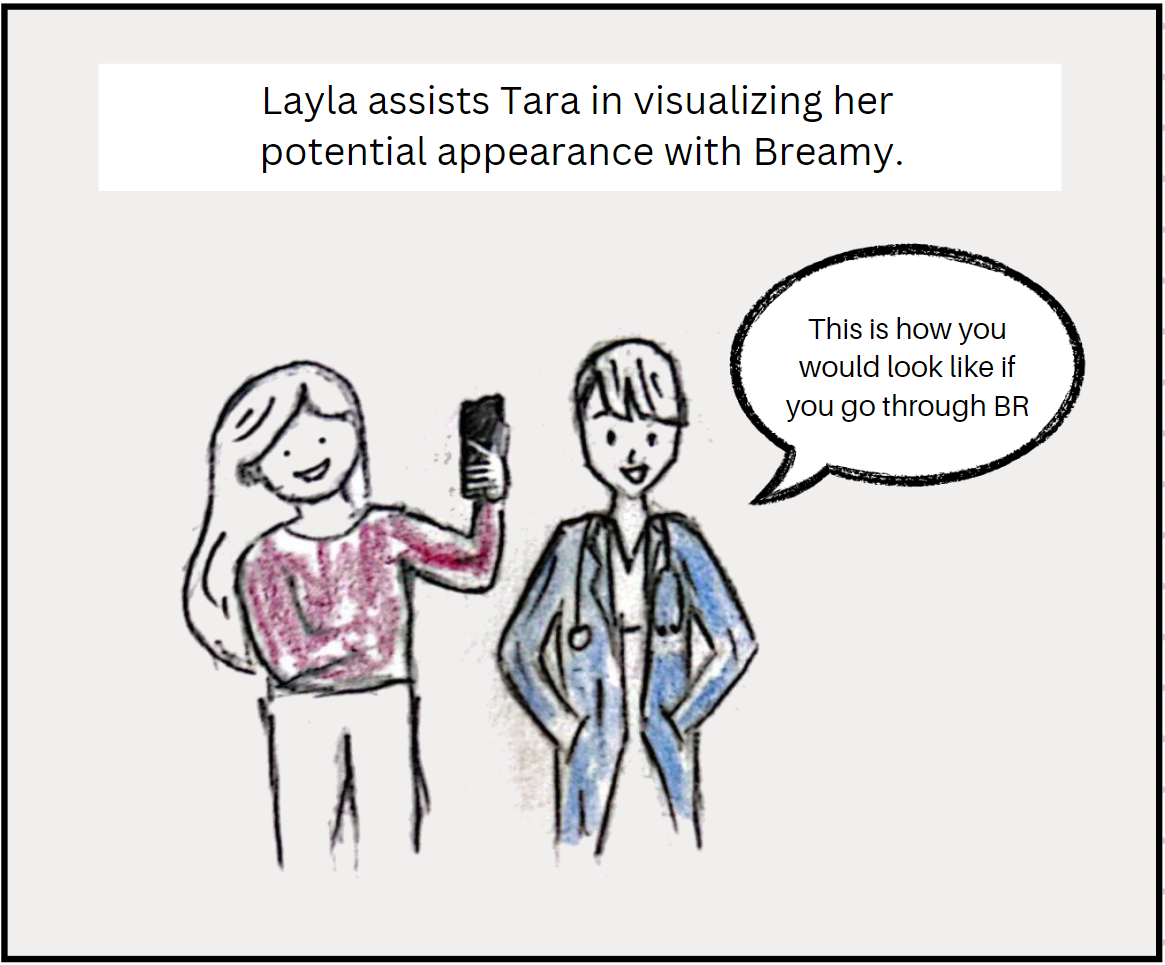}
    \caption{}
    \label{fig:3D visualization}
    \end{subfigure}
  \hfill
  \begin{subfigure}[b]{0.307\textwidth}
    \includegraphics[width=\textwidth]{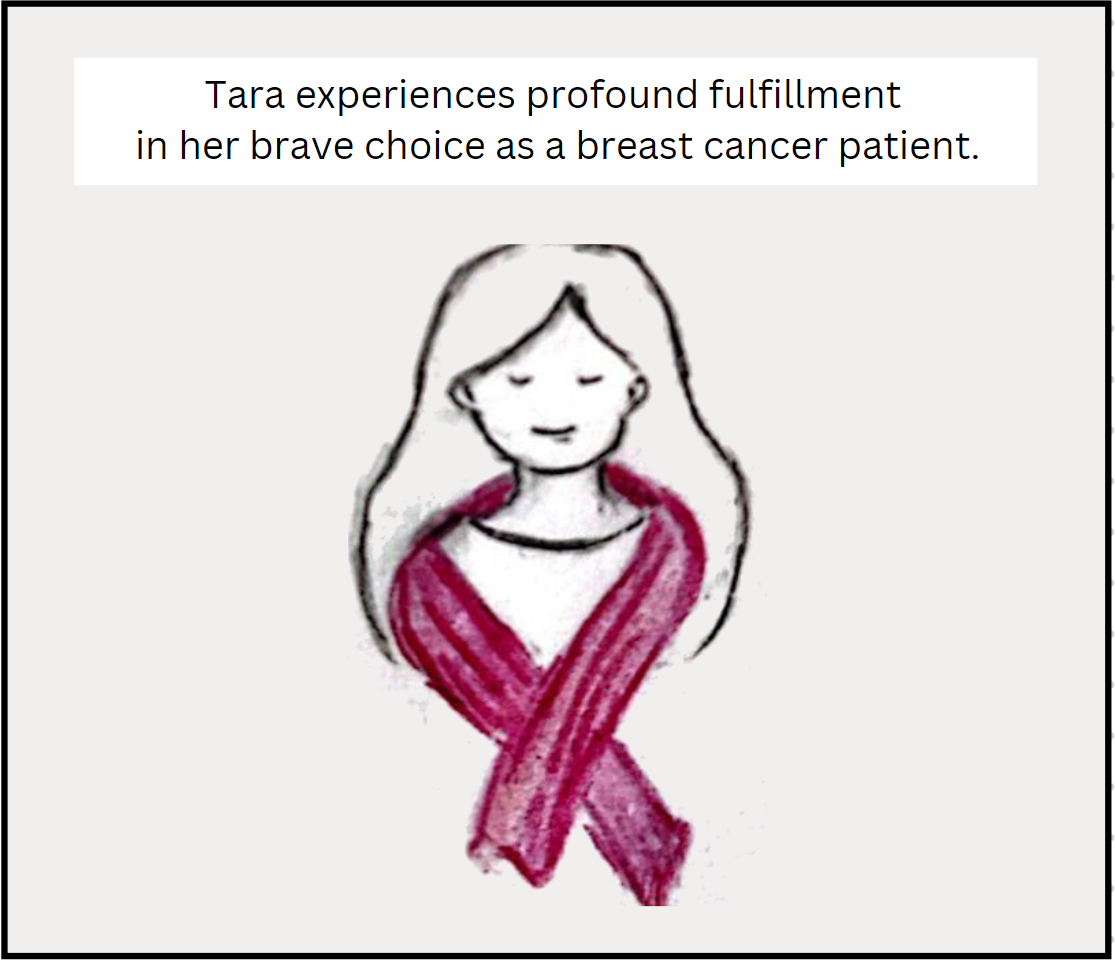}
    \caption{}
    \label{fig:3D visualization}
    \end{subfigure}
          \caption{Storyboard of how the surgical oncologist and patient interact with Breamy as a decision-making tool for 3D visualizing personalized oncoplastic options on patient's body.}
\label{fig:storyboard}
\end{figure}

As part of the design process, use case scenarios are often used to describe how a user might interact with a system to achieve a specific goals. We created a storyboard in which Tara, a fictional character has been recently diagnosed with breast cancer and narrate her experience and emotional engagement with her oncologist Dr. Laya (see Figure \ref{fig:storyboard}). In this fictional scenario, the surgeon discusses different oncoplastic options to help the patient feel more confident and less anxious in making their decision. After discussing several options with the surgeon, the various potential models and outcomes are saved in the patient's profile in Breamy. The patient then in the comfort of their home can use the AR module of Breamy. The patient uses their phone's camera and can swipe through the reconstruction and surgical options the surgeon has recommended and see how each will look on their body. 

\subsubsection{Breamy Modules}

\begin{figure}[h!]
  \centering
  \begin{subfigure}[b]{0.19\textwidth}
    \includegraphics[width=\textwidth]{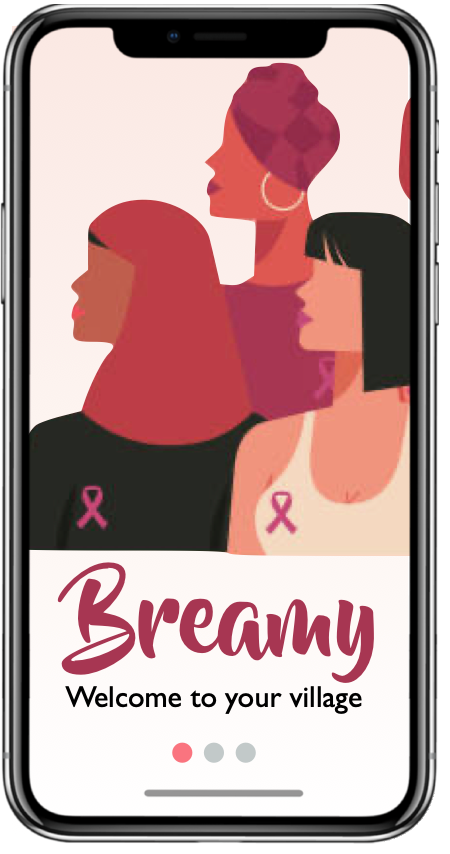}
    \caption{}
    \label{fig:image1}
  \end{subfigure}
  \hfill
  \begin{subfigure}[b]{0.192\textwidth}
    \includegraphics[width=\textwidth]{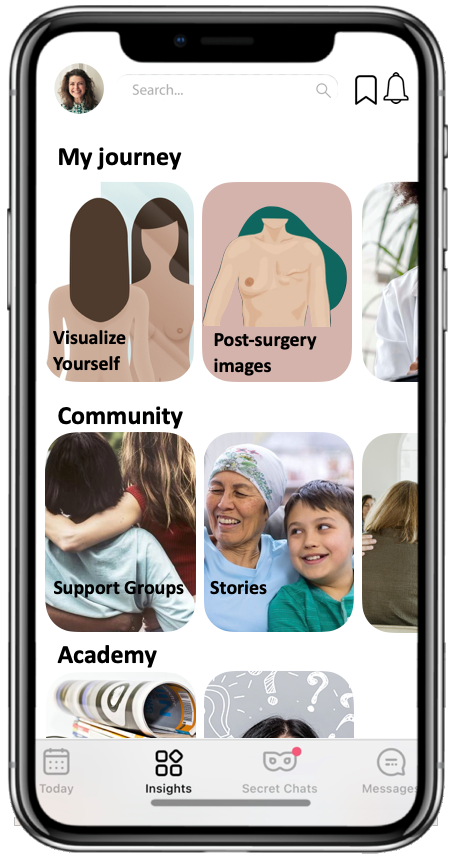}
 %   \caption{By exploring various oncoplastic options with AR, users will have the opportunity to visualize their potential appearance with or without clothing.}
 \caption{}
    \label{fig:3D visualization}
    \end{subfigure}
  \hfill
  \begin{subfigure}[b]{0.193\textwidth}
    \includegraphics[width=\textwidth]{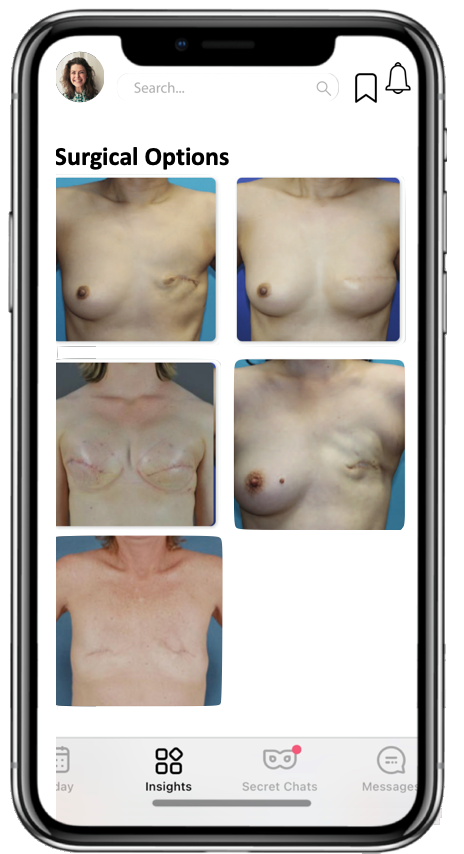}
    \caption{}
    \label{fig:3D visualization}
    \end{subfigure}
  \hfill
  \begin{subfigure}[b]{0.193\textwidth}
    \includegraphics[width=\textwidth]{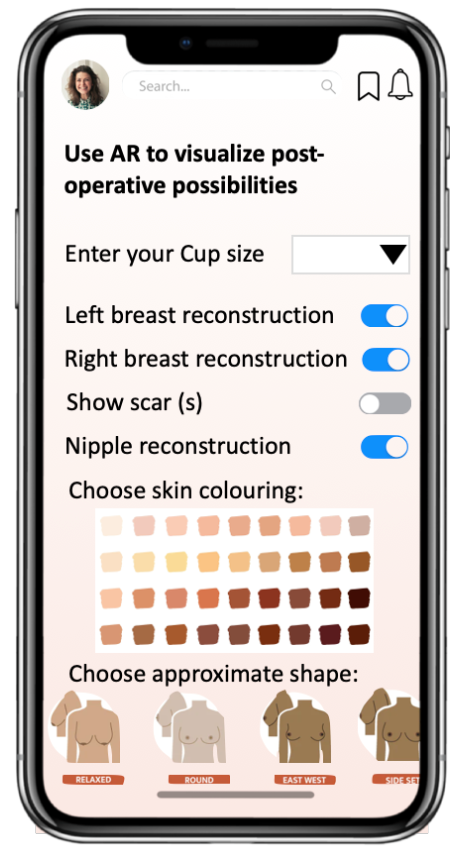}
    \caption{}
    \label{fig:3D visualization}
    \end{subfigure}
  \hfill
  \begin{subfigure}[b]{0.182\textwidth}
    \includegraphics[width=\textwidth]{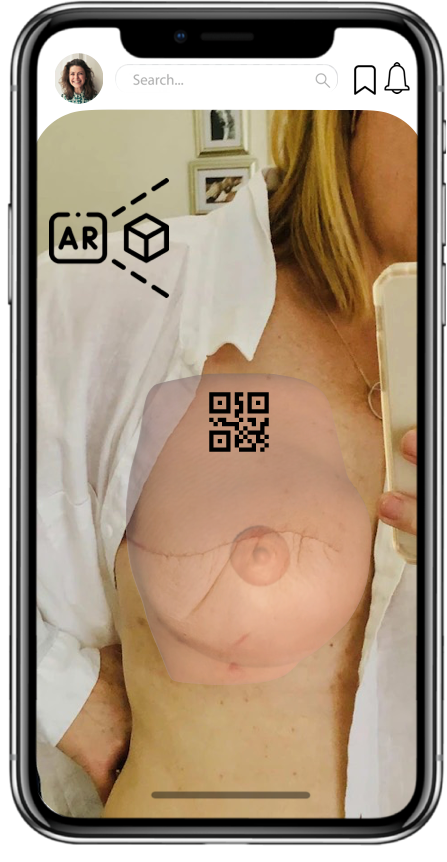}
    \caption{}
    \label{fig:3D visualization}
    \end{subfigure}
          \caption{Breamy UI design mock-ups: (a) onboarding screen (b) Homescreen consisting of all Breamy features (c) Post surgery oncoplastic procedures images (d) Filter 3D model based on patient's preferences (e) Projected customized 3D model on top of the QR-code with AR }
\label{fig:UI_design}
\end{figure}

Based on our research and analysis, we designed Breamy, to have an appealing interface, guidance through an AR feature that allow for visualizing various oncoplastic options, a knowledge repository to aid in informed decision making including information about post-surgery results, and a breast cancer community/support groups (see Figure \ref{fig:UI_design}). In this paper, we focus on the augmented reality decision-support feature which allows for personalized post-surgical oncoplastic options, however we briefly describe the reasoning behind each of the other modules we plan to implement of Breamy below.

\hfill

\noindent \textbf{Patient Journey:} Apart from seeking guidance from the surgical oncologist and plastic surgeon, patients often gather information from online sources such as the internet and social media platforms like YouTube \cite{tan2014patient}\cite{manne2016attitudes}. Furthermore, there has been shown to be a need for video-based information and accessible resources that can be viewed at home and shared with loved ones, particularly for those who travel long distances for their healthcare\cite{williams2017meeting}\cite{fristedt2021digi}. Therefore, in this module patients can browse through each oncoplastic option and view the images and video resources of similar procedure outcomes. Patients can save different 3D models/treatment options based on their own surgeons' recommendations. These models can be used in the augmented reality (AR) view to see how different oncoplastic procedures might look on their chests while they are at home. 
\newline

%According to the outcomes of our survey, individuals tend to initially consult healthcare providers and subsequently rely on online resources and websites to obtain information about breast cancer (see Figure \ref{fig:Breast_cancer_information_sources}). 

\noindent \textbf{Community:}
Writing and communicating on the Internet can have greater benefits for people with breast cancer compared to face-to-face discussions\cite{kim2012process}. Online platforms provide more time to think about and express responses to others' messages, promoting self-reflection and deeper communication. This can lead to improved health outcomes \cite{kim2012process}\cite{walther2002handbook}. For this reason, we will include online community groups to promote more meaningful communication. Furthermore, we plan to include patient stories as numerous previvors have expressed psychological gratification through the act of sharing information\cite{anderson2011uses}. By sharing their personal experiences, previvors can fulfill the need for a supportive community during their previvorship journey, which is not addressed by medical professionals \cite{wellman2022previvorship}.
\newline

\noindent \textbf{Academy:}
In a survey of 1000 women, it was found that nearly half of the participants were not aware of breast-conserving surgery and mistakenly believed that mastectomy was the only option\cite{sim2009breast}. Additionally, the study highlighted that the respondents had insufficient knowledge about the risk factors and held misconceptions about treatment and screening procedures. Therefore, Breamy aims to have the latest publications for users and little quizzes to test their knowledge.
\newline 

\noindent {\textbf{Augmented Reality:} The main goal behind the Breamy application was to develop a decisional aid for patients who are undergoing breast cancer surgery with oncoplastics. Thus we developed the AR module to allow patients to envision various surgical outcomes. In the first version of Breamy, we exclusively introduced breast reconstruction models. However, after initial feedback from cancer patients who had undergone mastectomy, we recognized a strong need to integrate deep inferior epigastric perforator procedure visualization (a kind of reconstruction that uses a woman's own tissue to develop a new breast after a mastectomy). Furthermore, patients mentioned how individual surgeons have varying techniques and thus it may be hard to accurately visualize surgeon specific outcomes. To meet these needs we aimed to create the 3D virtual models for decision making as follows. First, photographs of a surgeon's specific outcomes or photographs the surgeon believes are most likely to represent the patient's specific outcomes are chosen by the surgeon. These can then be mapped to a specific 3D model, either based on an implant shape or the patient's natural breast. The 3D models in our prototype were developed using Nomad and Blender software. Additionally, Photoshop was used to create 2D textures that were applied to the 3D models. The pipeline for the AR module is show in in Figure \ref{fig:3D_model_creation_procedure}. 

\begin{figure}[h]
    \centering
    \includegraphics[width=1\linewidth]{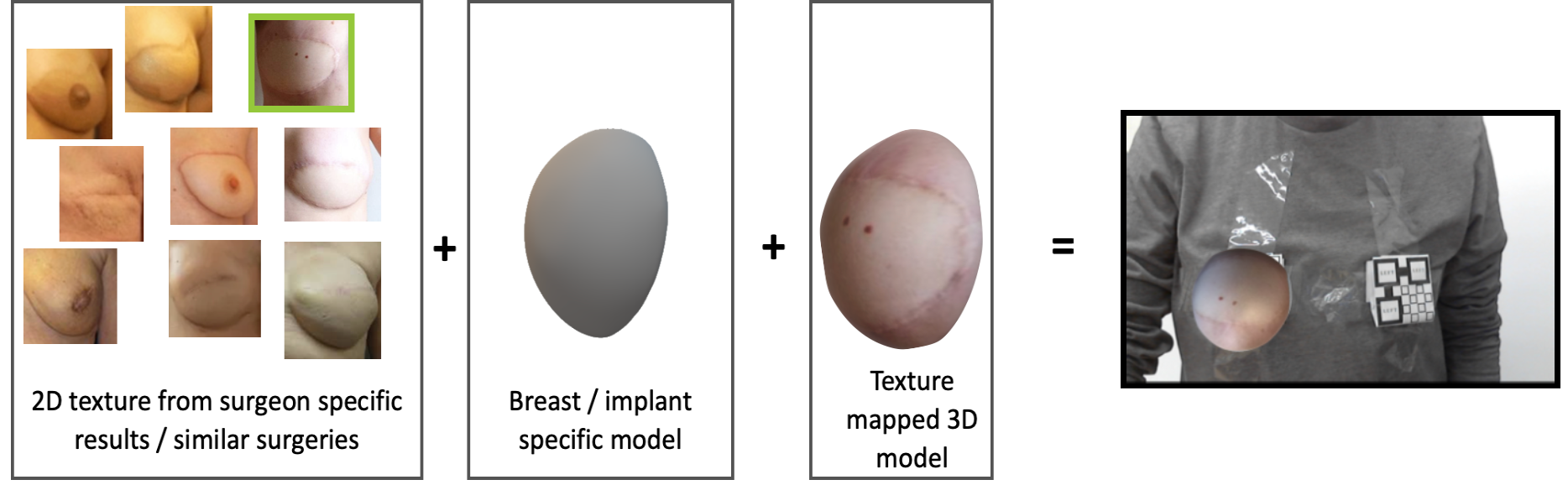}
    \caption{Pipeline going from 3D model creation to projection over patient's chest. From left to right: (1) the surgeon chooses from photographs previous surgical results which he believes are most similar to what the patient can expect, (2) specific breast/implant model(s) are chosen based on a discussion between the patient and surgeon, (3) the photograph texture is mapped to the 3D model, (4) the AR view is presented to the patient through the Breamy app.}   
    \label{fig:3D_model_creation_procedure}
\end{figure}

\begin{comment}
    
An early version of the Breamy application is shown in Figure~\ref{fig:3D_model_creation_procedure}. 
\begin{figure} [h!]
  \centering
  \begin{subfigure}[b]{0.5\textwidth}
    \includegraphics[width=\textwidth]{Bar_code_image.png}
    %\label{fig:image1}
  \end{subfigure}
  \hfill
  \begin{subfigure}[b]{0.4\textwidth}
    \includegraphics[width=\textwidth]{3D_visualization.JPG}
 %   \caption{By exploring various oncoplastic options with AR, users will have the opportunity to visualize their potential appearance with or without clothing.}
    %\label{fig:3D visualization}
    \end{subfigure}
     \label{fig:3D visualization}
    \caption{\empht{Left:} The yellow marks in the images show the extent to which the pattern recognition software AR can identify the pattern. \emph{Right:} By exploring various oncoplastic options with AR, users will have the opportunity to visualize their potential appearance with or without clothing.}
\end{figure}
\end{comment}

%\begin{figure}
 %\centering
 %   \includegraphics[width=1\linewidth]{capture.png}
 %   \caption{Sources for obtaining breast cancer information}
 %   \label{fig:Breast_cancer_information_sources}
%\end{figure}

\subsection{AR Module Implementation}
We developed the AR module to allow patients to envision various surgical outcomes. Breamy is currently designed for Android using C\# as the programming language, however, in future work we will port it to iOS. The AR module was developed using Unity’s AR Foundation framework and Vuforia Augmented Reality SDK. We have developed markerless AR within Breamy with the help of guideviews. To create this, we're utilizing Vuforia's Model Target Generator (MTG) software. MTG generates model targets by taking a 3D model that represents the object we want to track (in this case, a breast) as input. It assesses the model's suitability and allows us to configure it for the best tracking performance with a guideview. The MTG produces a vuforia database that we can employ with vuforia engine's Unity integration to enable object tracking. In addition to the marker-less method for automated recognition, we also all for a marker-based, where the patient places a tag on their chest. This method allows patients to see the surgical results while they are wearing their clothes for example when they are learning to use the app or in the context of a session with their surgeon or doctor.

It should be noted that there are no API calls made between Breamy and Vuforia servers. This means that all pattern recognition and 3D model projection processes happen without needing an internet connection and the 3D models can be projected instantly without storing any user data.

\subsubsection{Marker-based AR} For the marker-based AR view, we defined an image target for projecting the 3D model with respect to this marker. For the Vuforia engine to accurately detect and track our image target, it is important that the image target is detailed, has high contrast, and contains unique patterns \cite{vuforiaBestPractices}. The 3D model is projected upon detection of the AR marker, i.e., an image target that we have set as a QR code using the Vuforia Target Manager database for the project (see Figure \ref{fig:3D_model_creation_procedure}). Once this is detected, the user-selected surgical option is projected on the marker.

\subsubsection{Marker-less AR} For the markerless AR view, we use Vuforia's ``Model Targets" library. Model Targets enable apps to recognize and track objects in the real world based on their shape. To develop a Model Target for a particular object a 3D model data of that object is needed (e.g. CAD model or 3D scan). A Model Target requires users to position their device at a specific angle in relation to the object they want to track and maintain a particular distance to commence tracking. To facilitate this procedure, the application shows ``guideview", which estimates the representation of the object at the designated distance and viewing angle. By aligning a device with this image, the user initiates tracking as soon as the object aligns with guidelines. 

In our application, the Model Target should be the user's 3D breast shape that can then be used for recognizing and tracking the patient's breast. We use a guideview by presenting a rendered outline of a 3D model of the patient's breast. For the purpose of the prototype, we manually create the guideview (~Figure \ref{fig:guideview}) with the help of the MTG. The patient then adjusts their device until the guideline fits their torso to start tracking. In future work, we will use 2D images to 3D model to automatically create the 3D models and guideviews as described in the discussion section.

\begin{figure}[h!]
  \centering
  \begin{subfigure}[b]{0.49\textwidth}
    \includegraphics[width=\textwidth]{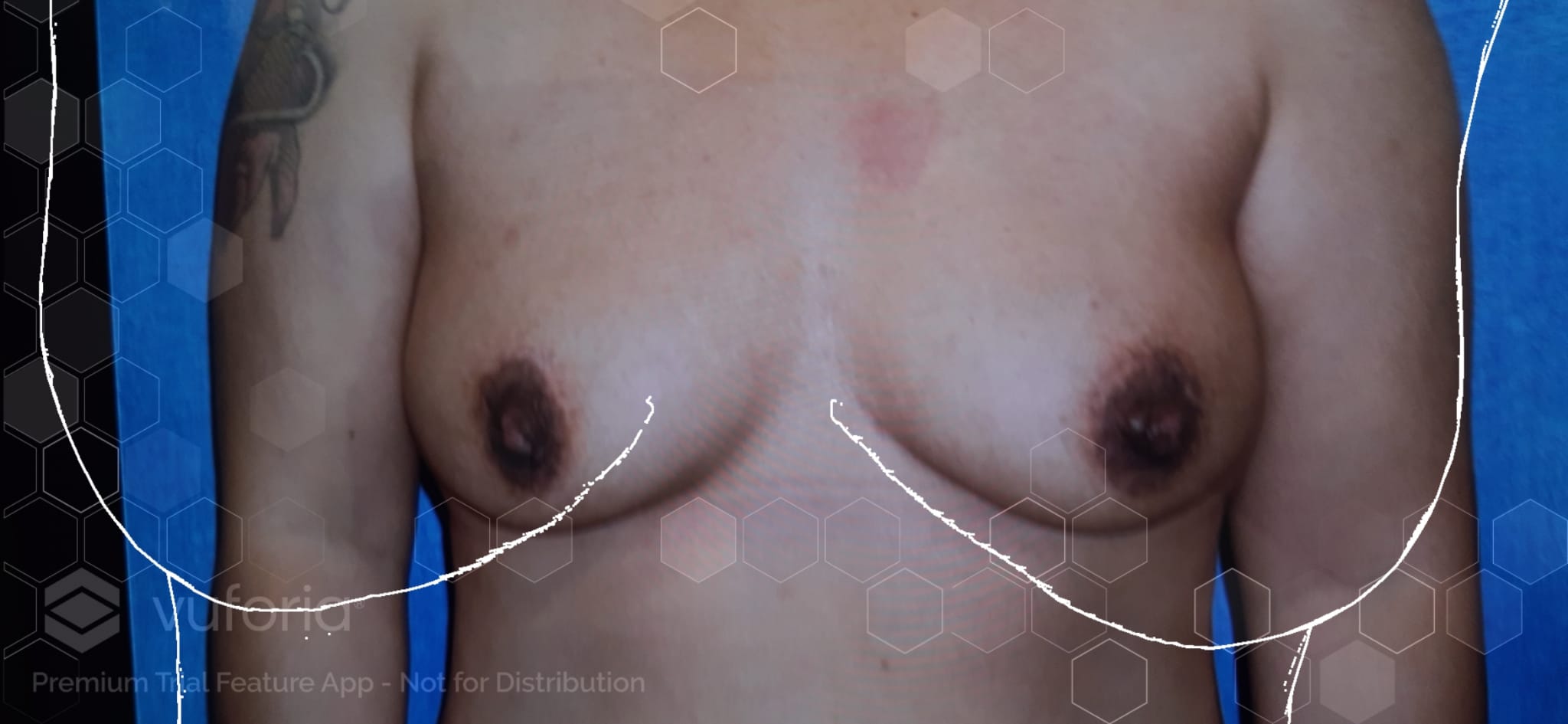}
    \caption{}
    \label{fig:guideview}
  \end{subfigure}  
  \begin{subfigure}[b]{0.49\textwidth}
    \includegraphics[width=\textwidth]{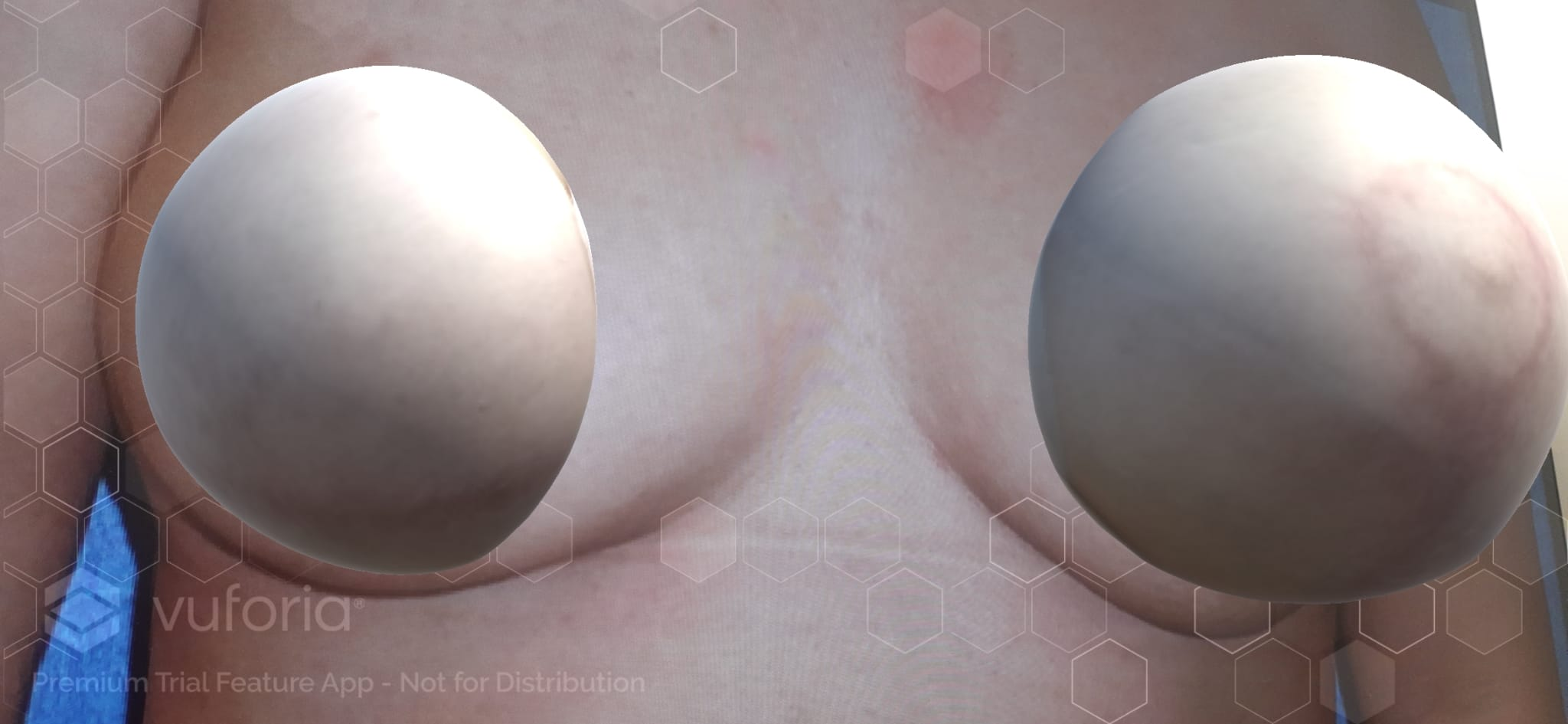}
    \caption{}
    \label{fig:3D visualization}
    \end{subfigure}
          \caption{ (a) Guideview rendered over the camera image. The user moves their body / camera until they fit the guideview. Once this is done the virtual models with possible oncoplastic options are projected onto the patients' breasts (b). }
\label{fig:guideview}
\end{figure}

\section{Evaluation}
To assess the design features, as well as, the perceived usefulness of Breamy we developed a survey including demographic, medical, and Breamy design feedback questions. Our main objective was to assess participants' knowledge of breast cancer and augmented reality (AR), as well as their perspective on the significance and usefulness of the Breamy AR application in the context of a breast cancer journey. We distributed the survey through social media platforms, including Facebook support groups, LinkedIn, Twitter, etc. to reach our participants and gather their valuable insights. The questions of the survey are shown in Table \ref{tab:Breamy Survey Questions}.

\begin{table}[h!]
%\captionsetup{font=small}
\caption{Breamy Survey Questions}
\label{tab:Breamy Survey Questions}
\begin{tabular}{>{\raggedright\arraybackslash}p{0.45\textwidth} >{\raggedright\arraybackslash}p{0.45\textwidth}}
\hline
\textbf{Demographic questions}      & \textbf{Breamy-related Questions}                         \\ \addlinespace\hline
\\
%\quad Under 18          & 0    {\color[HTML]{6e6d6d}  (0\%)}                       \\
 What sources of information do you typically rely on when seeking information about breast cancer?          & Have you used any other digital platforms or mobile applications related to breast cancer (screening, community, forums, etc.)?                      \\ \\ 
How important do you believe it is to have access to accurate and comprehensive information about preventative or breast cancer treatment options? &  How comfortable are you with viewing your body within an augmented reality (AR) application?
\\ \\
Have you ever undergone breast cancer screening (mammogram/ultrasound)?           & Have you used any augmented reality (AR) applications before?                        \\ \\
Are you aware of the different oncoplastic options available to breast cancer patients who undergo mastectomy (e.g., breast reconstruction, breast implants, nipple reconstruction, etc.)?            & How likely would you be to use an augmented reality (AR) application that showcases different oncoplastic treatment options projected onto your body if you had to make a decision about your own breast cancer treatment?                       \\ \\

How important do you believe oncoplastic are in the breast cancer journey or risk management process? &  Please specify any particular oncoplastic treatment options that you would like to see demonstrated in the augmented reality (AR) application.
\\ \\ 
What are the specific challenges or difficulties you have faced in accessing or understanding information about oncoplastic treatments (e.g., lack of awareness, etc.) &   Does the application have all the features that you have expected? Please specify any other feedback or suggestions you would like to provide regarding the augmented reality (AR) application or its potential impact. \\ \\ 
Would being able/having been able to visualize post-operative surgical results (on your own body) help you in treatment decision making? & How likely do you think an augmented reality (AR) application with personalized oncoplastic treatment information could improve patient comprehension and decision-making?                \\ \\
\hline
\end{tabular}
\end{table}
\hfill

\section{Results}
%\subsection{Participant Demographics}
Overall, Our survey included 166 participants, and it's important to note that not all participants answered every question. In fact, 135 participants completed the majority of the survey questions. Of the 166 participants that provided their age, 89\% were over the age of 35 years. Among the 135 participants that completed the entire survey, 65 participants (48\%) were breast cancer patients (48\%), 38 participants (28\%) were at high risk for breast cancer, and 32 participants (24\%) were at average risk for breast cancer. Furthermore, 162 out of 166 participants (97\%), were identified as women, and 4 were non-binary. The demographics of our survey participants are shown in Table \ref{tab:Respondents' demographics}. %\%Three-fourths of them were between the ages of 35 and 64, with the majority of them being women (97\%) and Caucasian (81\%). 
 %add use of digital platforms related to breast cancer percentages (and name them)

\begin{table}[h!]
%\captionsetup{font=small}
\caption{Survey Respondents' Demographics}
\label{tab:Respondents' demographics}
\begin{tabular}{p{0.5\textwidth} >{\raggedleft\arraybackslash}p{0.5\textwidth}}
\hline
\textbf{Demographics}      & \textbf{Total (\%)}                         \\ \addlinespace\hline
Age group (years) & {\color[HTML]{333333} (n = 165)} \\
%\quad Under 18          & 0    {\color[HTML]{6e6d6d}  (0\%)}                       \\
\quad 18 - 24           & 5    {\color[HTML]{6e6d6d}  (3\%)}                       \\
\quad 25 - 34           & 13   {\color[HTML]{6e6d6d}  (8\%)}                        \\
\quad 35 - 44           & 49   {\color[HTML]{6e6d6d}  (30\%)}                       \\
\quad 45 - 54           & 42   {\color[HTML]{6e6d6d}  (25\%)}                       \\
\quad 55 - 64           & 34   {\color[HTML]{6e6d6d}  (21\%)}                       \\
\quad 65 - 74           & 17   {\color[HTML]{6e6d6d}  (10\%)}                       \\
\quad 75 - 84           & 5    {\color[HTML]{6e6d6d}  (3\%)}                       \\
%\quad 85 or older       & 0    {\color[HTML]{6e6d6d}  (0\%)}                       \\
\addlinespace
Gender                    & (n = 166)                        \\
%\quad Male                      & 1 {\color[HTML]{6e6d6d}(1\%)}                          \\
\quad Woman                    & 162 {\color[HTML]{6e6d6d}(97\%)}                       \\
\quad Non-binary / Third gender & 4 {\color[HTML]{6e6d6d}(2\%)}                          \\
%\quad Other                     & 0 {\color[HTML]{6e6d6d}(0\%)}                          \\
%\quad Prefer not to say         & 0 {\color[HTML]{6e6d6d}(0\%)}                           \\

\addlinespace
Ethnicity                                    & (n = 165)                        \\
%\quad American Indian or Alaskan Native            & 0 {\color[HTML]{6e6d6d}(0\%)}                          \\
\quad Asian / Pacific Islander                     & 19 {\color[HTML]{6e6d6d}(12\%)}                         \\
\quad Black or African American                    & 1 {\color[HTML]{6e6d6d}(1\%)}                           \\
\quad Hispanic                                     & 3 {\color[HTML]{6e6d6d}(2\%)}                           \\
\quad White / Caucasian                            & 134 {\color[HTML]{6e6d6d}(81\%)}                        \\
\quad Multiple ethnicity/ Other  & 8 {\color[HTML]{6e6d6d}(5\%)}                             \\

\addlinespace
Education level                                                                                  & (n = 163)                        \\
%\quad No schooling completed                                                                           & 0 %{\color[HTML]{6e6d6d}(0\%)}                          \\
%\quad Nursery school to 8th grade                                                                      & 0 {\color[HTML]{6e6d6d}(0\%)}                          \\
\quad Some high school, no diploma                                                                     & 1 {\color[HTML]{6e6d6d}(1\%)}                          \\
\quad High school graduate, diploma or equivalent (e.g. GED) & 17 {\color[HTML]{6e6d6d}(10\%)}                        \\
\quad Trade/technical/vocational training                                                              & 11 {\color[HTML]{6e6d6d}(7\%)}                         \\
\quad Associate degree                                                                                 & 13 {\color[HTML]{6e6d6d}(8\%)}                         \\
\quad Bachelor’s degree                                                                                & 43 {\color[HTML]{6e6d6d}(26\%)}                        \\
\quad Master’s degree                                                                                  & 49 {\color[HTML]{6e6d6d}(30\%)}                        \\
\quad Professional degree                                                                              & 15 {\color[HTML]{6e6d6d}(9\%)}                         \\
\quad Doctorate degree                                                                                 & 14 {\color[HTML]{6e6d6d}(9\%)}                         \\
%\quad Other                                                                                            & 0 {\color[HTML]{6e6d6d}(0\%)}                             \\
\addlinespace
Diagnosis status                                    & (n = 135)                        \\
\quad Breast cancer patient            & 65 {\color[HTML]{6e6d6d}(48\%)}                          \\
\quad High risk individuals                     & 38 {\color[HTML]{6e6d6d}(28\%)}                         \\
\quad Average risk individuals                    & 32 {\color[HTML]{6e6d6d}(24\%)}                           \\
\hline
\end{tabular}
\end{table}
\hfill

\subsection{Patient-centered Decision-making}

In the survey, we had various clinical questions and asked about the role of technology as a decision aid in the breast cancer journey including participants' knowledge about different oncoplastic options and the potential of using augmented reality as a decision aid tool. We found that around half of the breast cancer patients (61 patients in total) lacked the necessary knowledge about the various treatment options (such as chemotherapy, mastectomy, etc.) available to them during their breast cancer journey. Additionally, around 15\% of the breast cancer patients expressed their dissatisfaction regarding the provided information or held neutral opinions. In the context of knowledge about various surgical options, we found around three-fourths of high-risk and breast cancer patients (103 out of 135 participants) were aware of the various oncoplastic alternatives offered to breast cancer patients having mastectomy. On the other hand and not surprisingly, only 13\% of the average risk participants were aware of oncoplastic options (4 out of 32 people). Based on the 135 responses, we also found that a significant majority, around 80\% recognize the importance of oncoplastic surgery and 10\% held a neutral stance regarding its significance in the breast cancer journey or in the risk management process.

All participants believed that the accessibility to accurate and comprehensive information about breast cancer treatment options is important. Out of 60 patients with breast cancer, almost 20\% reported dissatisfaction with their experience accessing information, particularly regarding oncoplastic treatments. We also asked breast cancer patients about any challenges they faced with receiving treatment information. Out of 43 comments from breast cancer patients, several mentioned feeling pressured to decide quickly without easy access to reliable or accessible information to help them make informed decisions. This suggests a strong need for better access to surgical information and more decisional support. 

%-- satisifaction of treatment options\newline 

\subsection{Breamy Perceptions}
 Approximately half of breast cancer patients (27 out of 55 patients) currently utilize digital platforms or mHealth applications. Similarly, this percentage extends to about 30 to 40 percent of individuals within the high-risk and average-risk population categories respectively. These platforms include Facebook breast cancer support groups and various applications that enable health monitoring and provide other health-related features. Out of a total of 112 respondents, almost 90\% had never used an AR application. However, 90\% and more of the individuals believed that an augmented reality (application with personalized oncoplastic treatment information could improve patient comprehension and decision-making (see Table \ref{tab:participant-views}). Moreover, many breast cancer patients mentioned that when they felt pressured and had limited access to information, an AR decision aid tool would significantly help them make decisions.

\begin{table}[h!]
\caption{Participants' views on the potential of improving patient comprehension and decision-making through the usage of AR application visualizing different oncoplastic treatment information.}
\label{tab:participant-views}
\begin{tabular*}{\textwidth}{@{\extracolsep{\fill}}llll@{}}
\hline
\textbf{Diagnosis Status}      & \textbf{Unlikely to improve(\%)}                    & \textbf{Neutral or likely to improve(\%)}               & \textbf{Total} \\ \hline
Breast cancer patient &     5     {\color[HTML]{6e6d6d} 10\%} & 44 {\color[HTML]{6e6d6d} 89\%}    & 49    \\
High risk individual  & 1   {\color[HTML]{6e6d6d} 3\%}                        & 27 {\color[HTML]     {6e6d6d} 96\%}                           & 28    \\
Average risk individual    & 1 {\color[HTML]{6e6d6d} 3\%}                     & 31 {\color[HTML]{6e6d6d} 97\%}                        & 32    \\
\end{tabular*}

\end{table}

 Out of 61 breast cancer patients (and 55 that responded to all questions), 49 of them stated that visualizing post-operative surgical results on their own body would aid in treatment decision-making. Across all participants, approximately 80 percent (out of 110 people) expressed their willingness to use Breamy to showcase different oncoplastic treatment options projected onto their body if they had to make a decision about their own breast surgery treatment (see Table \ref{tab:willingness rates to use Breamy AR application}).

\begin{table}[h!]
\caption{Willingness to use Breamy AR application}
\label{tab:willingness rates to use Breamy AR application}
\begin{tabular*}{\textwidth}{@{\extracolsep{\fill}}llll}
\hline
\textbf{Diagnosis Status}                             & \textbf{Unwilling(\%)} & \textbf{Neutral or Willing(\%)} & \textbf{Total} \\ \hline
{\color[HTML]{333333} Breast cancer patient} & 12 {\color[HTML]{6e6d6d}(24\%)}     & 38 {\color[HTML]{6e6d6d}(76\%)}              & 50    \\
High risk individual                         & 3 {\color[HTML]{6e6d6d}(11\%)}      & 25 {\color[HTML]{6e6d6d}(89\%)}              & 28    \\
Average risk individual                       & 6 {\color[HTML]{6e6d6d}(19\%)}      & 26 {\color[HTML]{6e6d6d}(81\%)}              & 32    \\ \hline
\end{tabular*}
\end{table}

 Positive comments from participants included: "Sounds fantastic! I would've liked to have this when I had my  [Photobiomodulation therapy], and probably in the future with revisions too", "This seems like a very positive step in helping those who need to make these decisions feel a bit more comfortable and have a better understanding of what options they have, and how they might look", and "I think it's a great idea. I use google images before to see what AFC [Aesthetic Flat Closure] would look like and I have shared my photo across the flat unreconstructed site so other women would know what to expect".  
 
 Comments from participants also included a desire for additional options like nipple reconstruction and adding tattoos in the AR module. Numerous individuals also expressed a desire to envision their appearance going flat (removing the breasts and having a flat chest), and visualizing post-operative scars but achieving this solely through the use of AR technology is challenging. While augmented reality enhances reality, it is more difficult to eliminate or subtract from it. 
 
 A small number of participants were more negative about the utility of the application specifically because it did not allow visualization of flat or Goldilocks procedures (i.e. removing the breast tissue but leaving a layer of fat under the skin and using the deep layer of the skin from the lower part of type breast to fold inside to make a breast mound). Three participants also felt the app may not be able to show realistic enough reconstruction results due to variations in surgical techniques or other considerations besides aesthetics (e.g. vessels and blood flow) which might be detrimental and cause misguided expectations. This will be an important aspect to consider in future work. 
 
 A number of participants who had been diagnosed with breast cancer also suggested additional features that would be useful in the application including helping woman understand the need for therapy after surgery, adding links to counsellors and therapists, and providing a discussion of risks and complications for various types of surgery.

\section{Discussion}
The results from the literature review, needs assessment, and survey distributed among 166 participants indicate the necessity for providing a tool for aiding patients in deciding their oncoplastic treatment option. Our preliminary findings indicate that augmented reality has the potential to help patients visualize various surgical options in order to make a more confident decision. Furthermore, the 3D visualization can be used to improve communication during the preoperative planning stage of surgery and to provide more personalized healthcare for patients. 

The majority of survey participants showed keen interest in visualizing the DIEP procedure, which uses a woman's own tissue to reconstruct a new breast after a mastectomy. Participants also would like to see post-mastectomy information such as exercises and how the procedure's recovery would look like. We aim to include potential complications and how they would affect the patient's appearance and recovery times within the application for each option to help patients better decide and set their expectations. Furthermore, as pointed out by a few of our participants realism would have a big impact on the usefulness of our application. Therefore, in future work, we plan to work on the visualization of the AR to improve realism. We will not only add virtual elements in the form of surgical results but also filter the camera image, e.g. adding blur and transparency so that the virtual element does not seem to float above the image. Another option would be to add a snapshot feature that would allow us to do offline processing which are a set of actions performed independently from the real-time AR interaction. During this phase, adjustments, enhancements, or mappings are applied to the user's skin tone and texture to better align them with the 3D models. 

In future work, to personalize this process for each patient, the application will use photogrammetry techniques. This involves capturing multiple images of the patients' breasts from various angles while keeping the viewpoint change within a 30-degree range to create the model target \cite{vuforiaModelTargets}. This will allow us to create patient-specific 3D models that can be sent to the PTC Vuforia servers through the model target web API for a deep-learning-based training process. This process will generate an advanced Model Target dataset, extending recognition ranges up to 360 degrees. %In advanced model targets the recognition range in which we want the model target to be detected is specified in degrees. %In our case we want the breast to be detected in XX degrees(Figure X).%
Unlike standard model targets that rely on guideviews, advanced model targets enable the object to be recognized and tracked from any position within the specified recognition range, eliminating the need for users to manually align the camera view with the physical object. We plan to test the application with cancer patients as well as broaden our investigation by analyzing the influence of potential complications and the expertise of surgeons on the visual outcomes of oncoplastic procedures.  Finally, we intend to conduct a clinical trial to assess the degree of similarity between the anticipated reconstruction outcomes and the actual results.

\section{Conclusions}
In breast cancer, 3D visualization/simulation has proven to be an effective implant selection tool and a method of aiding patients in making decisions \cite{lee2013time, crivellari2000burdens}. The aim of this study was to introduce Breamy, an accessible AR decision aid application for breast cancer patients undergoing mastectomy. A survey of 166 participants showed that 90\% of the participants believed Breamy would be effective as a decision aid. The results of this preliminary work indicate that using AR as a decision-aid tool for breast cancer patients could enhance patient understanding and assist patients in making informed decisions. 

%In order to perform this study, we developed an AR application with the Unity software containing four main parts: a predefined pattern for 3D visualization of the oncoplastic treatment option, an image repository for viewing specified procedures outcomes following the oncoplastic surgery, online community groups for promoting meaningful communication, and a comprehensive repository of the latest publications and resources to increase breast cancer knowledge. 

\nocite{*}% Show all bib entries - both cited and uncited; comment this line to view only cited bib entries;

\begin{comment}
    \bmsection*{Author Biography}

\begin{biography}{\includegraphics[width=76pt,height=76pt,draft]{empty}}{
{\textbf{Author Name.} Please check with the journal's author guidelines whether
author biographies are required. They are usually only included for
review-type articles, and typically require photos and brief
biographies for each author.}}
\end{biography}
\end{comment}

\end{document}